\documentstyle[aps,prc,preprint,tighten]{revtex}

\newcommand{\TR}{\mbox{$T_{\rm R}$}}

\begin{document}
\draft
\title{An inversion procedure for coupled-channel scattering:\\ 
determining the deuteron-nucleus tensor interaction.}

\author{S.G. Cooper$^{\dag}$, V.I. Kukulin$^{\ddag}$, R.S. Mackintosh$^{\dag}$
 and  V.N. Pomerantsev$^{\ddag}$}
\address{$^{\dag}$Physics Department, The Open University, Milton Keynes,
 MK7 6AA, U.K.\\$^{\ddag}$Institute of Nuclear Physics, Moscow State 
University, Moscow 119899, Russia. }

\date{\today}
\maketitle
\begin{abstract}
We present a practical $S$-matrix to potential inversion procedure
for coupled-channel scattering. The 
inversion technique developed is applied to  
non-diagonal $S^J_{ll'}$ for spin one projectiles,
yielding a tensor interaction \TR\, and is also 
applicable to spin-$\frac{1}{2}$ plus spin-$\frac{1}{2}$ scattering. 
The method is a generalization of the iterative-perturbative, IP, method. 
It is tested and evaluated and we investigate the
degree of uniqueness of the potential, particularly for  cases
where there is insufficient information to define the potential uniquely.
We examine the potentials which result when the $S$-matrix is
generated from a $T_{\rm P}$ interaction.  
We also develop the generalisation, using established procedures,
of IP $S$-matrix-to-potential inversion
 to direct observable-to-potential inversion. 
This `direct inversion' procedure is demonstrated to be an efficient 
method for finding  a multi-component 
potential including a \TR\ interaction
fitting multi-energy $\sigma$, ${\rm i}T_{11}$, $T_{20}$, $T_{21}$ and $T_{22}$ 
data for the scattering of spin-1 nuclei from spin-zero target.
It is applicable to other channel spin 1 cases.

\end{abstract}

\pagebreak

\setlength{\parindent}{0.3 in}

\section{Introduction}
Various methods for carrying out $S$-matrix to potential inversion are now 
available, see for example~\cite{chadan,balaton}, but, until recently, it has 
been possible only for cases with channel-spin  zero or 1/2. However, 
there have been many accurate experiments involving 
spin-one polarised particles and these provide a powerful motivation to develop an 
efficient technique for inversion in cases with higher channel 
spin, i.e. coupled-channel scattering. With such a technique, one can 
exploit the very large volume of polarisation data which has accumulated.  
This includes vector and tensor analysing powers and polarisation transfer 
observables.

Over the years, 
we have developed a practical and widely generalisable procedure, the iterative 
perturbative, IP, method~\cite{early,kalinin,ketal1,zuev,candm} and we  
recently~\cite{prl} 
demonstrated an extension to spin-1 projectiles for the first time. 
Ref.~\cite{prl} demonstrates the method in one specific application, 
but does not give details or a derivation of the method.
In this paper we present details of our extension of the IP $S$-matrix to potential
inversion method to the coupled-channel case of spin-1 projectiles and present
further evaluation of it. We also test and evaluate an important extension of 
the IP method,   single step data-to-potential inversion for coupled-channel 
scattering.

In Ref.~\cite{prl} we applied single step inversion to analyse real data.
We note here that there are also many ways in which IP 
$S\rightarrow V$ inversion can contribute to understanding
nucleus-nucleus interactions. Perhaps the most obvious application 
is the inversion of $S$-matrix elements found by phase shift analysis 
of experimental data. However, there are also many important applications
which involve the inversion of theoretical $S$-matrix elements, i.e.\
 elastic channel $S$ derived from 
coupled channel calculations, Glauber model and resonating group model 
calculations. The potential found in this way
contains information concerning the contribution of tensor components to 
dynamic polarization and 
exchange processes to inter-nuclear potentials. For the particular case 
of spin-polarised deuteron and $^6$Li scattering, 
obvious applications include the study of the influence 
of reaction channels and distortion effects on the projectile-nucleus 
potential, especially in its non-central components. There  
are a number of longstanding puzzles relating to spin-polarised deuteron scattering, 
including the anomalously small real part of the tensor interaction,
which can be studied using these methods.

Coupled-channel inversion represents a significant
development in inversion techniques since  a non-diagonal potential
is derived from  a non-diagonal $S$-matrix. Such a potential couples channels of
the same conserved quantum numbers but different values of orbital angular 
momentum.
Spin-1 inversion is therefore the first example of coupled channel inversion 
in which a non-diagonal $S$-matrix is made to yield a non-diagonal potential.
Apart from the general derivation, the
present paper  is framed for a very specific two-channel case:
deuterons scattering from a spin zero nucleus with inversion
determining a  tensor interaction involving the
non-diagonal operator $T_{\rm R}$ as given below.
In spite of the rather specific nature of the present
application, we believe this work opens the way to a fully general
class of coupled-channel inversion situations involving the determination
of a coupling potential from a non-diagonal $S$-matrix.

This paper presents in detail only those aspects of the IP inversion
formalism which are connected
with the specific coupled channel generalisation to spin-1 scattering. 
In other respects it calls upon previous publications~\cite{ketal1,zuev,candm}
in which the general aspects of the  IP inversion procedure are described. 
Because of the specific application to spin-1
projectiles, we establish our notation by beginning in Section II with a brief 
review of basic aspects of spin-1 scattering. 

An important feature of the IP $S\rightarrow V$ inversion procedure  
is the natural way in which it can be convoluted with $
{\rm (data)} \rightarrow S$ fitting to give 
an overall  ${\rm (data)} \rightarrow V$ algorithm.
This provides a new and efficient data analysis tool  
which in many cases obviates the need for
independent ${\rm (data)} \rightarrow S$ inversion. Important advantages follow
when fitting data for many energies since the underlying potential model guarantees
that the energy dependence of the $S$-matrix will be smooth without the need
to postulate parameterized forms for $S(E)$. Indeed, we have shown~\cite{prl58}
that  ${\rm (data)} \rightarrow V$ inversion can provide a powerful alternative method
for phase shift fitting of multi-energy scattering data for light nuclei. 

Ref.~\cite{prl} contained a restricted analysis 
of low energy $\vec{\rm d}$ + $^4$He data. A future paper will present a much 
more exhaustive analysis of the very large
collection of data for this system. At a later stage we hope to present
an analysis of  $^6\vec{\rm Li} + ^4$He  data, including tensor analysing powers.

\section{The scattering of spin-1 nuclei from spin-0 targets}

\subsection{Formalism for spin-1 scattering}
In order to establish our notation,
we outline the standard formalism for the elastic scattering of spin-1
projectiles from a spin-0 target in the presence of tensor forces. 
 The key feature introduced by the tensor interaction of \TR\ type (see below
for a classification of tensor forces) is that it couples channels of different
orbital angular momentum $l$. Specifically, for particular values
of the conserved quantities  $J$, the   total angular
momentum,  and $\pi$, the parity, whenever $\pi = (-1)^{J+1}$,  then two  
values of orbital angular momentum, $l=J-1$ and $J+1$, are  coupled by \TR. 

For total angular momentum $J$ and orbital angular momentum $l'$ the 
radial wavefunction $\psi^J_{l'l}$ satisfies the coupled equations,
\begin{equation}
\left[ \frac{{\rm d}^2}{{\rm d} r^2} + k^2 -\frac{2\mu}{\hbar^2}(l'J|V|l'J) 
-\frac{l'(l'+1)}{r^2}
\right] \psi^J_{l'l}(k,r) 
= \sum_{l'' \ne l'} \frac{2\mu}{\hbar^2}(l'J|V|l''J)  \psi^J_{l''l}(k,r)
\label{eq:cde}
\end{equation}
where $\mu$ is the reduced mass of the system and $(l'J|V|l''J)$, a function of $r$, 
is the matrix element
of the inter-nuclear interaction $V$ integrated
over all angular and internal degrees of freedom. The second subscript, $l$, 
on $\psi$ identifies
the incoming orbital angular momentum. This is determined by imposing 
on the solution of the coupled equations, the following
asymptotic boundary conditions :
\begin{equation}
\psi^J_{l'l}(k,r) \rightarrow \delta_{l'l} I_{l'}(kr) - S^J_{l'l} O_{l'}(kr).
\label{eq:boundary}
\end{equation} 
Here, $I_l(r)$ and $O_l(r)$ are the incoming and outgoing asymptotic
Coulomb radial wavefunctions, often  written $H_l(r)^{*}$ and $H_l(r)$ respectively 
as in Satchler~\cite{satchlerbook}, namely:
$$ I_l(kr)=G_l(kr)-{\rm i} F_l(kr);\qquad I_l(kr)=G_l(kr)-{\rm i} F_l(kr);$$
where $F_l$ and $G_l$ are regular and irregular Coulomb wavefunctions 
respectively. Note that the boundary conditions given 
in Eq.~\ref{eq:boundary}
differ by a factor from those adopted by Satchler.
Where there is no ambiguity, we suppress the $J$ superscript. When
$\pi= (-1)^J$,  $V$ is diagonal and Eq.~\ref{eq:cde} is uncoupled.

In general, $S$ will not be unitary, but will be subject to the unitarity limits:
$|S_{11}|^2 + |S_{12}|^2 \le 1$ and $|S_{22}|^2 + |S_{21}|^2 \le 1$, where, of course 
$S_{12}=S_{21}$. These limits present no particular problem for $S \rightarrow V$
inversion where $S$ can be assumed to satisfy them, but they can
represent a significant problem in the case of data to potential inversion, 
see Section~\ref{datatp}.

\subsection{The $T_{\rm R}$ interaction and its effect.}\label{matrixele}
Non-diagonal matrix elements $(l'J|V|l''J)$  occur in Eqn.~\ref{eq:cde} for 
elastic scattering of spin-1 projectiles
with certain types of tensor force.  The possible forms of local tensor 
interaction have been classified by Satchler~\cite{satchler,satchlerbook} 
who defined \TR, $T_{\rm L}$ 
and $T_{\rm P}$ interactions. The $T_{\rm L}$ interaction is 
believed~\cite{stamp} to be very small, at least below 50 MeV/u,
and is in any case diagonal in $l$. The $T_{\rm P}$ interaction could well be
substantial~\cite{andy} but appears to be hard to distinguish phenomenologically
from \TR. The gradient operators within  $T_{\rm P}$ make calculations harder, and
the present inversion method does not apply to it.
\subsubsection{The \TR\ operator} In this work  we assume 
that the inter-nucleus potential $V$ contains a tensor force component 
of \TR\ form~\cite{satchlerbook}: 
\begin{equation} \TR V_R(r) \equiv ({\bf (s\cdot \hat{r})^2} -2/3) V_R(r).
\label{eq:tensor}
 \end{equation}

We quote the matrix elements of the  interaction \TR\ for future reference.
 The diagonal matrix elements of \TR\ are:
\begin{center}
\begin{tabular}{||c|c|c|c||} \hline
 & $l=J-1$ & $l=J$ & $l=J+1$ \\ \hline
$<Jl|\TR|Jl>$ & $-\frac{1}{3} \frac{J-1}{2J+1}$ &
$ \frac{1}{3}$ &$ -\frac{1}{3}\frac{J+2}{2J+1}$\\
\hline \end{tabular}
\end{center}
and  the non-diagonal matrix elements are:
\begin{equation} <J\, J-1|\TR|J\, J+1> = <J\, J+1|\TR|J\, J-1> = 
\frac{[J(J+1)]^{1/2}}{2J+1} \label{eq:diagTR}. \end{equation}

\subsubsection{The radial form of the \TR\ interaction}
The derivation of $ V_R(r)$ from the folding model has been discussed at length
long ago by Keaton and his colleagues~\cite{kaprc8,kav,keaton} and also
by Raynal~\cite{raynalpl}. Within the folding model, the deuteron
\TR\ interaction arises directly from
the D-state component. The overall general success of folding models 
for central and spin-orbit interactions suggests
that folding model calculations of $ V_R(r)$  should give at least approximately 
the correct radial form and overall magnitude, but this has not been borne out
in the case of \TR\ according to extensive phenomenological studies, 
e.g.\cite{frickprl,clementzphys,frickzphys,matoba,ermer}.
The overall conclusion is that the real part of $ V_R(r)$  predicted by the folding 
model is much too strong for heavy nuclei, of the right order of magnitude for light 
target nuclei, and actually about three times too strong for a $^4$He 
target.\footnote{Ref~\cite{prl} 
exploits the inversion formalism presented here to give an alternative analysis
of d + $^4$He scattering, a theme elaborated in later papers.} These facts, 
together with a large literature discussing breakup and 
reaction channel contributions,
suggest that we have no generally applicable reliable knowledge of $V_R(r)$.
There is reason to doubt even the general arguments, based on folding models,
 that it should be small in the interior of heavier nuclei, 
away from the nuclear density gradients in the surface. Such gradients define the 
angle between the projectile spin ${\bf s}$ and the vectorial position
${\bf r}$ of the projectile with respect to the nuclear centre, 
see Eq.~\ref{eq:tensor}.

\section{$S \rightarrow V$ inversion for spin-1 projectiles on spin-0 targets}
\subsection{General background of IP inversion}
The IP method has been  successful
for $S_{lj} \rightarrow V(r) + {\bf l \cdot s} V_{\rm so}(r)$ inversion for
spin half projectiles, and we now present its generalisation to spin-1 inversion. 
The only restriction is to  a \TR\ tensor interaction. 
Certain features of the IP method, to our knowledge not shared by other inversion 
methods,  will be of particular importance in the particular systems to which we 
shall apply spin-1 inversion. These include the ability to find an explicitly
energy dependent potential from phase shifts for a range of energies, the ability to 
handle a range of energies simultaneously and to include Majorana terms
for all potential components.  For many applications the important property is that
mentioned in the introduction, i.e.\ that IP inversion lends itself to 
direct observable to potential inversion. 
This not only avoids the need for independently determined phase
shifts (or $S$-matrix), but actually provides an advantageous method of 
determining such phase shifts. For a full description of IP
inversion as applied in the spin-1/2 case  see Refs.\cite{ketal1,zuev,candm}. 
The formalism presented in Ref.\cite{candm}, whereby energy dependent potentials  
are obtained
from multi-energy datasets, can be used with spin-1 inversion as described here,
although energy dependence is not actually exploited in the test cases.
A brief general account of IP inversion is given in the next section.

\subsection{IP inversion for the coupled channel case; 
application to spin-1}\label{ipinv}
Our notation must reflect the fact that
the outcome of inversion will be a potential with many components.
We therefore label each component with an index $p$ which identifies central,
spin-orbit or tensor terms, each real or imaginary. The number of components
doubles when the potential is parity dependent. (Parity dependence
is particularly important for light nuclei at lower energies.)

The IP method commences with a `starting reference potential', SRP, and proceeds
by iteratively correcting each component $p$ of the potential:
\begin{equation}
V^{(p)} \rightarrow V^{(p)} + \sum_n \alpha^{(p)}_n v^{(p)}_n(r) \label{first}
\end{equation}
where $\alpha^{(p)}_n$ are coefficients to be determined and $v^{(p)}_n(r)$
are the  functions comprising the `inversion basis',
 (which, if required, can be chosen differently for different $p$). 
The amplitudes $\alpha^{(p)}_n$ are determined at each iteration from 
linear equations, based on an SVD algorithm,
which successively reduce the `phase shift distance' $\sigma$ defined by:
\begin{equation}
\sigma^2 = \sum |S^{\rm t}_k - S^{\rm c}_k|^2 \label{sigma}.
\end{equation} For each partial wave $k$, $S^{\rm t}_k$ is the `target' 
$S$-matrix and $S^{\rm c}_k$ is for the potential at the current iteration. 
Here the label $k$ is a single index which identifies the partial wave 
angular momentum $l$ as well as the energy $E_i$ when multi energy sets of 
$S_l(E_i)$ are simultaneously inverted. It also includes
non-diagonal elements of $S^J_{ll'}$ in the spin-1 case described later.

The linear equations are based on the (usually) very linear
response~\cite{ketal1,daresbury}, $\Delta S$, of the complex $S$-matrix to small 
changes $\Delta V$ in the potential. The expression for this is well 
known  in the uncoupled case and is very simple:
\begin{equation}
\Delta S_l = \frac{{\rm i} m }{\hbar^2 k}
\int_0^{\infty} (\psi_l(r))^2 \Delta V(r) {\rm d}r.
\label{second}
\end{equation}
In Eq.\ref{second}, the $S$-matrix $S_l$ is written in terms of the asymptotic
form of the regular radial wave function as $\psi_l(r) 
\rightarrow I_l(r) - S_l O_l(r)$
where $I_l$ and $O_l$ are incoming and outgoing Coulomb wave functions
as before. When inverting $S_l(E_k)$ over a series of energies $E_k$,
the energy label $E_k$ is implicit in these equations with index $k$ subsumed 
with orbital angular momentum $l$ to give an overall channel label. In the case 
of spin-1/2, the $j$ label is also subsumed in the same way~\cite{candm}.
Linear algebraic equations for local variations of $\alpha^{(p)}_n $ follow 
from the minimisation of $\sigma^2$,~\cite{early,kalinin,ketal1}.

We now present the generalized linear response relationship which applies
to the non-diagonal $S$-matrix for spin-1 elastic scattering. The derivation
is given in Section~\ref{derivation}. For any given set of conserved quantum 
numbers, certain channels will be coupled by the nucleus-nucleus interaction 
and we use labels $\kappa, \lambda, \mu, \nu$ for these channels. Thus the 
matrix element of the nucleus-nucleus interaction $V$ between the wavefunctions 
for channels $\kappa$ and $\lambda$, corresponding to integrating
over all coordinates but $r$, will be written $V_{\kappa\lambda}(r)$.
The increment $\Delta S_{\kappa\lambda}$ in the non-diagonal S-matrix which is 
due to a small perturbation $\Delta V_{\kappa\lambda}(r)$,  is\begin{equation}
 \Delta S_{\kappa\lambda} = \frac{{\rm i}\mu}{\hbar^2 k} 
 \sum_{\mu\nu}\int_0^{\infty}\psi_{\mu\kappa}(r) \Delta V_{\mu\nu}(r)
\psi_{\nu\lambda} {\rm d}r \label{integ}\end{equation} 
where $\psi_{\nu\kappa}$ is the $\nu$th channel (first index) component of 
that coupled channel solution for the unperturbed non-diagonal potential
for which there is in-going flux in channel $\kappa$ (second index) only.
The normalisation is
$\psi_{\nu\kappa} \rightarrow \delta_{\kappa\nu} I_{l_{\kappa}} - S_{\nu\kappa} 
O_{l_{\nu}}$ where
$I_l$ and $O_l$ are incoming and outgoing Coulomb wavefunctions for
orbital angular momentum $l$; there is
no complex conjugation in the integral. Starting from Eq.\ref{integ}, spin-one
inversion becomes a straightforward generalisation of the procedure 
outlined above and  described
in Refs.~\cite{ketal1,zuev,candm}. The method is implemented in the code 
IMAGO~\cite{imago} where the linearity relations have been exhaustively 
tested by the gradient method.

A convenient feature of the IP method is that one can judge from the behaviour
of $\sigma^2$ as the iteration proceeds whether a satisfactory inversion has
been achieved. A low value of $\sigma^2$ obviously guarantees that a potential
closely reproducing the input $S_l$ has been found. Because the IP method is
implemented interactively, there is an opportunity to examine the potential 
for oscillatory features. These might well be spurious and
result from over-fitting noisy data.  In such a case, one can reduce the
basis dimensionality or raise the SVD limit and this generally allows one to
achieve a smooth potential, often with only a small increase in $\sigma^2$.
One must bear in mind that genuine oscillatory features, corresponding  to
non-locality in an $L$-independent local potential or to $L$-dependence of 
the underlying potential, can be necessary to achieve a precise representation
of $S_{lj}$ or $S^J_{l'l}$. 

We stress here that, although the context of our discussion is the 
determination of a tensor interaction from the non-diagonal $S$-matrix elements
describing the scattering of spin-1 projectiles, the range of application is much 
more general. 
 
\subsubsection{Derivation of non-diagonal perturbation expression.}\label{derivation}
We now outline the derivation of Eq.\ref{integ}. 
 The derivation  can be applied to the general coupled 
channel inversion from non-diagonal $S$-matrix to non-diagonal potential.
Our starting point is Eq.\ref{eq:cde} which
we shall write with a simplified notation for two channels. Until the last step
in the argument, we shall assume we are using units in which $\hbar^2/2m =1$.

The radial wavefunctions in channel $i$ with incoming waves in channel $\lambda$
are written as $\psi_{i\lambda}$ and have asymptotic behaviour 
at $r\to\infty$:
\begin{equation}\psi_{i\lambda} \rightarrow \delta_{i\lambda}I_{\lambda} - S_{\lambda i}
O_{i}\end{equation}
where for simplicity we write $O_i$ for the outgoing Coulomb wavefunction 
with orbital
angular momentum $\ell$ appropriate to channel $i$, and similarly for the 
ingoing wavefunction $I_{i}$. For brevity  
we omit labels for conserved quantum numbers $J$ and $\pi$.

Absorbing the centrifugal interaction within the potential, 
we can write the coupled equations
for the radial wavefunctions appropriate to incoming waves 
in channel $\lambda$ as:

\begin{equation} \psi_{i\lambda}'' = 
\sum_j (V_{ij} -E\delta_{ij})\psi_{j\lambda}, \qquad i=1,2.
\label{eq:one}\end{equation}
In the case considered here, $V_{ij}$ with $i\ne j$ arises entirely from 
the tensor interaction, the matrix $V_{ij}$ being symmetric.
Now, denoting by $\bar{\psi}$ the wavefunction arising from a (symmetric) 
perturbation in the potential $V_{ij}\rightarrow V_{ij} + \Delta V_{ij}$, 
we can write:
\begin{equation} \bar{\psi}_{i\lambda}'' = \sum_j (V_{ij} +\Delta V_{ij} 
-E\delta_{ij})\bar{\psi}_{j\lambda}, \qquad i=1,2.
 \label{eq:two}\end{equation}
Multiplying Eq.\ref{eq:one} by $\bar{\psi}_{i\mu}$ and Eq.\ref{eq:two} by
 $\psi_{i\mu}$, summing over $i$ and then subtracting the second one from the 
 first one, we get: 
\begin{equation} \sum_i \frac{\rm d}{{\rm d}r}(\psi_{i\lambda}' 
\bar{\psi}_{i\mu} -\bar{\psi}_{i\lambda}'\psi_{i\mu}) = 
-\sum_{ij} \psi_{i\mu}\Delta V_{ij} \bar{\psi}_{j\lambda},
\label{eq:three} \end{equation}
as all terms including $V_{ij}$   in the right hand side vanish due 
to symmetry of $V_{ij}$.

Integrating  Eq.(\ref{eq:three}) from $r=0$ to the 
asymptotic region and using the usual Wronskian relationship
$W[I_{\ell},O_{\ell}] = -2 {\rm i} k$, we get:
\begin{equation} 
 2 {\rm i} k (\bar{S}_{\mu\lambda} - S_{\lambda\mu})=
 \sum_{ij}\int_0^{\infty}\psi_{i \mu} \Delta V_{ij} \bar{\psi}_{j \lambda} 
 {\rm d} r. 
\label{eq:four} \end{equation}
Since the coupled channel equations with a symmetrical potential matrix give 
a symmetrical $S$-matrix $S_{\lambda\mu} = S_{\mu\lambda} $, 
the left hand side of Eq.(\ref{eq:four}) is equal 
\begin{equation}
 2 {\rm i} k (\bar{S}_{\mu\lambda} - S_{\mu\lambda})= 
  2 {\rm i} k \Delta S_{\mu\lambda}
\end{equation}  
Hence we find, reinstating the $2m/\hbar^2$ factor on the right hand side:
\begin{equation}
\Delta S_{\lambda \mu} = \frac{{\rm i} m}{\hbar^2 k} \sum_{ij}\int_0^{\infty}
\psi_{i \lambda} \Delta V_{ij} \bar{\psi}_{j \mu} {\rm d} r. 
\label{eq:delS}\end{equation}
This expression is valid for any symmetrical finite perturbations 
$\Delta V_{ij}$ 
decreasing at $r\to\infty$ sufficiently rapidly. It is easily extended to 
any system of coupled-channel equations. For small $\Delta V_{ij}$ we make 
the Born approximation assumption that $\bar{\psi} \sim \psi$ and get:
\begin{equation}
\Delta S_{\lambda \mu} = \frac{{\rm i} m}{\hbar^2 k} \sum_{ij}\int_0^{\infty}
\psi_{i \lambda} \Delta V_{ij} \psi_{j \mu} {\rm d} r. 
\label{eq:dS}\end{equation}
The expression (\ref{eq:dS}) is the basis for the coupled-channel inversion 
method. The success of
the inversion method in leading to a converged solution confirms the wide 
applicability of this equation in each step of our iteration process.

\section{Testing couple-channel $S \rightarrow V$ inversion for spin-1 
projectiles}
We carried out two contrasting tests of $S \rightarrow V$ inversion as described 
in Sections~\ref{firsttest} and~\ref{multi} below.
First let us  define the potentials and the inversion basis. 

\subsection{Specification of the interaction potential and basis used}\label{specs}
In one respect our notation is non-standard.
We write down the complete potential for spin-1 projectiles scattering from
a spin-zero target. It is
\begin{equation} V_{\rm cen}(r) + {\rm i} W_{\rm cen}(r) + V_{\rm coul}(r) + \\
2{\bf l \cdot s}(V_{\rm so} + {\rm i}W_{\rm so}) + (V_{\rm R} + 
{\rm i}W_{\rm R})T_{\rm R}
\label{eq:defpot}\end{equation} where  $V_{\rm coul}(r)$ is the usual hard-sphere 
Coulomb potential. Note that 
our spin-orbit potentials $ V_{\rm so}$ and  $W_{\rm so}$, 
are defined in such a way that
they will be half the magnitude of those defined according to the usual 
convention~\cite{satchlerbook} for  spin-1 projectile.\footnote{In our papers 
relating to spin 1/2 projectiles, the usual convention has been used.} 
For the test cases we present, the spin-orbit potential is 
defined as in Eq.~\ref{eq:defpot}.

For notational simplicity, Eq.~\ref{eq:defpot} has not 
been written to reflect parity dependence. There are two alternative methods of
representing parity dependence. The code IMAGO can apply either of these to
each of the components in Eq.~\ref{eq:defpot} except $V_{\rm coul}$. The first
representation  defines Wigner and Majorana components for each term, say $V_{\rm x}$:
\begin{equation} V_{\rm x} = V_{\rm x, W} + (-1)^l V_{\rm x, M}\end{equation}
where $l$ is the partial wave angular momentum. With this form
the inversion procedure can be made to determine $V_{\rm x, W}$ and $V_{\rm x, M}$ 
for any or all $V_{\rm x}$. An alternative approach is to determine independent 
positive or negative parity components for $V_{\rm x}$. In many cases, the 
Wigner-Majorana representation is most natural and has been shown~\cite{kuk-new} 
to be preferable where the odd parity term may otherwise be ill-determined.
However, sometimes the odd-even representation is more appropriate, for example
where a particular $V_{\rm x}$ has completely different shapes and magnitudes 
for the different parities, as we believe can be the case for $V_{\rm R}$.  
The code IMAGO offers the  freedom to represent the parity dependence of 
each component $V_{\rm x}$ in either way.

The IP method is not tied to any particular set of functions for the
inversion basis and each component of the potential can be represented
by a different basis. It is an important
feature of the IP method, as implemented in IMAGO, that a range of different
functions is available, and those which we have applied are specified in~\cite{imago}.
Zeroth order Bessel functions and harmonic oscillator
functions are both linearly independent sets which have proven useful where
bases of large dimensionality are necessary to describe a potential over a wide
radial range down to $r=0$. For cases involving light nuclei, particularly for inversion
of small $S$-matrix datasets, a small basis comprising a series of Gaussian functions 
is preferable. A Gaussian basis covering just the nuclear surface
region is also useful for heavy ion cases where
there is no information available to determine the potential in the nuclear
interior. It is important that a  basis should not be chosen which would describe
the potential over a radial range, or to a degree of detail, which is not warranted
by the information contained in the set \{$S_l$\} or by the nature of the physical
situation.  In practice, much smaller bases 
are often necessary in order to eliminate spurious oscillatory features from
the potentials. The operation of the SVD algorithm, with adjustable
SVD limit, stabilises the inversion and can, where appropriate, reduce the effective 
dimensionality of the  inversion basis.

\subsection{Single-energy inversion}\label{firsttest}

In Ref.~\cite{prl} we presented a test of $S \rightarrow V$ for deuterons scattering
from the light nucleus $^4$He in which there is very little absorption. Here we present
a test for a much heavier nucleus, where there is substantial absorption, and
demonstrate that a potential, very accurate almost to the nuclear centre,
can be obtained by inversion. The test case
studied was for a $^{58}$Ni target and deuterons at a laboratory energy of 56 MeV. 
The parameters found by Hatanaka {\em et al\/}~\cite{hatanaka} fitting angular 
distributions and the analyzing powers $A_y$ and $A_{yy}$ were used.
A notable feature of the potential was that the imaginary $T_{\rm R}$ term was
quite large, but the real $T_{\rm R}$ term was very small (a common but unexplained
feature of deuteron optical potentials.) The spin-orbit component was real.
 The potential was parity independent as expected for
this combination of target, projectile and energy.
The energy and other characteristics of the reaction are such that there are
`many' active partial waves. `Many' here means sufficient, even 
with $S^J_{l'l}$ for a single energy,
to yield  a precise reproduction of the potential.

The test was carried out as follows: one of the authors applied the optical model 
parameters  of Hatanaka {\em et al\/} to
the standard spin-1 scattering code DDTP~\cite{ddtp}, reading out the 
$S$-matrix onto a file. A second author,
knowing only the target and the energy, then applied IMAGO to find the potential
from these $S$-matrix elements. The inversion was carried out with a starting 
potential which contained only real and imaginary central potential 
components. These were guessed from general systematics without specific knowledge
of Hatanaka's potential.  Since with IMAGO there is complete  
freedom to choose the  starting potential and inversion basis, it is worthwhile
to test the inversion method starting with no more information about the 
potential than might be available in a `for real' case.    

When a converged solution was found, the potentials obtained were compared with 
the known potentials with results shown in Figure 1. 
The solid lines represent the `target'
potential, i.e.\ the potential from which the $S$-matrix was calculated using
DDTP. The short dashes represent the potential found by inversion;  it can be seen
to reproduce the target potential very closely except very near to the nuclear
centre in the case of the imaginary tensor component. 
We have not shown the real tensor component which was only a few percent of 
the imaginary component nearly everywhere. This small component was reproduced only
qualitatively, as expected, since the absolute errors for the real and imaginary
$T_{\rm R}$ components were similar in magnitude and comparable to the
real $T_{\rm R}$ potential itself. In Figure 1, the dashed
line represents the starting potential, zero for the spin-orbit and tensor
components. The $S$-matrix elements for the target and inverted potentials
are indistinguishable on a graph, corresponding to values of $\sigma$ of 
roughly $10^{-3}$. 

This test shows that the inversion procedure has the capability of revealing 
quite fine details of the potential as would be required for the kind of 
applications,  discussed in the introduction requiring  the inversion of single 
energy $S$ derived from theory. Such studies might establish, for example, 
the contribution of specific exchange
terms or reaction couplings to the inter-nucleus potential.
 
\subsection{Multi-energy inversion at very low energy}\label{multi}

At low energies and for light target nuclei, very few partial waves are 
involved so that there will in general be insufficient information 
contained in the  $S$-matrix elements for a single energy to yield a 
detailed and precise potential. The situation  is even worse in cases where
parity dependence must be assumed since this halves the information available 
for potential components of each parity. The problem can be ameliorated if $S$ 
is available for more than one energy. If $S$ is available over a narrow range
of energies, then  the algorithm can be made to yield to an energy independent
potential; this is what we have earlier called `mixed case' inversion (see 
Refs.~\cite{ketal1,zuev} and first of Ref.~\cite{candm}) and, in effect, 
the information from the energy 
derivative of $S$ is exploited. In many cases,  $S_{lj}$ or $S^J_{l'l}$ 
are  provided over  a wide range of energies. In this case one should ideally
consider the potential to be energy dependent and determine the energy 
dependence itself. This can be done within the framework of the 
parameterisations presented above. 

An example of where the sets of $S_{lj}$ or $S^J_{l'l}$ are too small to
define the potential very closely is the
$S\rightarrow V$ inversion situation embedded in the analysis
of low energy, experimentally determined, multi-energy observables
for d + alpha scattering. A first report was presented in Ref.~\cite{prl}.
The  test we now describe is directly relevant and
asks the following question:  what properties of the potential can reliably 
be determined from very small sets of $S$?  

The test was for deuterons scattering from $^4$He with $S^J_{l'l}$ calculated 
from a known potential at 11 energies: 8, 8.5 \ldots 12.5, 13 MeV.
The known potential was energy independent but parity dependent and  
was taken to be real. (The imaginary parts of empirical potentials 
are known to be small for d + $^4$He at these energies.) The following terms 
were included: central Wigner, central Majorana, spin-orbit Wigner
 and separate even parity and odd parity $T_{\rm R}$ tensor potentials
(the odd/even choice for $T_{\rm R}$ reflects what we believe~\cite{prl} 
to be the case for the actual d + $^4$He tensor force.) The central and 
spin-orbit terms are like those found in Ref.~\cite{prl}, and the 
very large even parity tensor term is based on that of 
Dubovichenko~\cite{dubo98}, see also \cite{kuk-new}.

The inversion was effectively `mixed case' in the sense just described.
The starting potential was zero in all components except for the Wigner
real central and Wigner real spin-orbit terms. In keeping with the nature
of this test, the  very small inversion basis of Ref.\cite{prl} was
used. This has two Gaussians only for 
each component except the central components for which there were three.
The centres and widths of the Gaussians were not varied during the inversion. 

The `target' (known) and inverted potentials are shown in Figure 2, together
with the starting potential required by the IP method. The starting 
potential is the dot dashed line, non-zero for two components only, and 
corresponding to $\sigma = 10.552$ where $\sigma$, defined in 
Section~\ref{ipinv}, is summed over the 11 energies. The inverted potential 
is shown as the dotted line, and the `target' potential, from  which 
$S^J_{l'l}$ was calculated, is the full line. We see that the qualitative 
features are reproduced although less well for the small components 
and near the nuclear centre. The value of $\sigma$ for the potential
shown in the dotted line was $0.135$ which is reasonable for 
a low energy multi-energy case. For 10.5 MeV, this corresponds to
$S^J_{l'l}$ for the target and inverted potentials being indistinguishable
on a graph apart from one single term: the phase angle of the non-diagonal 
part of $S^J_{l'l}$ for higher partial waves for which, in any case, 
the magnitude $|S^J_{l'l}|,\,\, l\ne l',\,$ is very small. 
The tensor potential, having very different odd and even parity components, 
is as well reproduced as could be expected with the very small basis. Note that
the starting potential for the inversion has zero tensor terms. From
the matrix elements of $T_{\rm R}$ given in Section~\ref{matrixele}, 
we see that $l=0$ partial waves are ineffective and hence we cannot expect
to reproduce the tensor real term $V_{\rm R}$ at $r=0$.

In summary: we found that the qualitative properties of the potential were 
reliably reproduced, particularly for the larger
components. Thus, reliable statements about the general features
of d + $^4$He potentials can be made, but nothing can be asserted concerning
non-central interactions for $r < 0.5$ fm. 

\section{Inverting $S^J_{l'l}$ calculated with a $T_{\rm P}$ tensor interaction.}
The inversion technique which we have described is limited to a tensor 
force of the $T_{\rm R}$ type. Since there exist processes which
are expected to lead to $T_{\rm P}$ forces, the possibility must be faced
that data analysed using the data-to-$V$ 
extension of the inversion method, which is described
in Section~\ref{datatp}, will indeed involve a $T_{\rm P}$ tensor interaction. 
It is therefore relevant to ask, in the context of
$S\rightarrow V$ inversion: can we invert $S^J_{l'l}$ calculated 
with a $T_{\rm P}$ tensor interaction with a potential 
which has only a $T_{\rm R}$ tensor interaction? If so, to what extent does
inversion yield  valid central and spin-orbit components? 

There is further interest in knowing how well the general effects of a  
$T_{\rm P}$ interaction can be represented by a 
$T_{\rm R}$ potential. The properties and even existence of a  
$T_{\rm P}$ interaction have not yet been convincingly linked to experiment
since the consequences of the two kinds of interaction
are difficult to distinguish phenomenologically. This was  discussed
by Goddard~\cite{goddard} who compared $S^J_{l'l}$ and the 
observables calculated from a  $T_{\rm R}$ interaction with the corresponding
quantities calculated from a particular $T_{\rm P}$ interaction 
 devised in such a way that, according to semi-classical 
arguments, it would  be very similar in effect.

We study these questions by exploiting
the equivalent pairs of tensor potentials introduced by Goddard.
We first inverted  $S^J_{l'l}$ for 30 MeV deuterons scattered from $^{56}$Fe 
with a $T_{\rm R}$ potential and then inverted  $S^J_{l'l}$ derived from
the potential containing that  
$T_{\rm P}$ interaction which is `equivalent' in Goddard's sense. The two 
potentials are given in Table 1 of Ref.~\cite{goddard}. 

The first part of the  test showed that inversion of  $S^J_{l'l}$ for a 
known $T_{\rm R}$ still works very well at about half the energy of the 
test described in Section~\ref{firsttest}. The results were very similar: 
the $T_{\rm R}$ potential, which in this case is of a volume Woods-Saxon 
form with depth 5 MeV, is accurately reproduced even at the nuclear centre. 
The solid and (scarcely distinguishable) dashed lines in Figure 3 respectively
represent Goddard's original potential and that found by inversion.
The $S^J_{l'l}$ for the inverted potentials, including the non-diagonal 
terms, are indistinguishable on a graph from those for the original potentials. 

The dotted lines in Figure 3, show the inversion for Goddard's $T_{\rm P}$ 
case. The  non-tensor components are qualitatively well reproduced, the derived 
potentials having  the appearance of the target potentials but with 
superimposed oscillations. This waviness is relatively  more significant
for the small components, the real central potential being  reproduced
to within a few percent for all $r$. The $T_{\rm R}$ interaction found by 
inversion is now surface peaked in form but of
average depth comparable to that of the Woods-Saxon  (which however had a 
local momentum dependence, see~\cite{goddard}). The {\em diagonal\/} 
$S^J_{l'l}$ for target and inverted potentials are graphically indistinguishable,
as are $\arg S^J_{l'l}$ for $l\ne l'$ for low values of $J$. However 
the non-diagonal $S$-matrix was not well reproduced for $J>7$, for which partial 
waves the non-diagonal $|S^J_{l'l}|$ is very small. The value of $\sigma$ 
was much higher than for the $T_{\rm R}$ case, i.e.\ 0.0294 compared with
0.00589. 

The results presented graphically in Figure 3 can be quantified in terms
of the volume integrals and rms radii for the central and spin-orbit
components of the inverted potentials. For the $T_{\rm R}$ case, all of these
quantities were reproduced to a few parts in a thousand with the (small)
volume integral of the spin-orbit term being least accurate: the error was
0.7 \%. The errors for the non-tensor components found when inverting
Goddard's $T_{\rm P}$ potential were a few percent, the volume integral 
of the spin-orbit term again being least accurate with an error of 3.8 \%.  

Goddard also performed an identical comparison for the case of 13.0 MeV 
deuterons scattering from $^{46}$Ti, and we repeated the test just described
for this case. There is interest in doing this since the inversion algorithm
applied to $ S^J_{l'l}$ for a single energy is expected to  fail
at lower energies for reasons explained in Section~\ref{multi}. 
However, we find that the
 results for 13 MeV deuterons on  $^{46}$Ti are essentially the same as
for 30 MeV deuterons on $^{56}$Fe for both $T_{\rm R}$ and $T_{\rm P}$ 
interactions. The form of the $T_{\rm R}$ potential representing the 
actual $T_{\rm P}$ component was essentially the same as that shown for 
30 MeV in the bottom panel of Figure 3 and this similarity applies also to
the deviations of the non-tensor terms. It therefore appears that we have
found  general properties of the $T_{\rm R}$ potential representing an 
actual $T_{\rm P}$  potential. 

As a result of these tests, and noting that  $T_{\rm P}$ interactions are 
not predicted to be particularly large, we conclude:
\begin{enumerate} \item The existence of
processes of the kind which give rise to a $T_{\rm P}$ component will not
prevent this inversion procedure, which includes only $T_{\rm R}$ tensor 
interactions, from fitting $S^J_{l'l}$ and is unlikely to greatly falsify 
inversions of this kind, particularly with regard to the non-tensor components. 
IP spin-1 inversion as described here is thus not fatally undermined by the 
possible existence of $T_{\rm P}$ interactions. The effort 
needed to develop  spin-1 inversion including $T_{\rm P}$ interactions 
require greater motivation than exists at present.
\item  As Goddard  suggested, almost all the effects of such a potential
can be well represented by a $T_{\rm R}$ tensor interaction, although its 
relationship to the form of the $T_{\rm P}$ interaction is, as might be 
expected, more complicated than can be deduced from simple semi-classical 
arguments~\cite{goddard}. The phenomenological problem of establishing  
$T_{\rm P}$ interactions is still considerable.
\end{enumerate}

\section{Data to potential inversion for spin-1 projectiles}\label{datatp}

In what follows, we first briefly indicate how $S\rightarrow V$ inversion
is extended to ${\rm (data)} \rightarrow V$ inversion for 
the uncoupled case, then indicate how this is extended to
include coupling, as is required for spin-1 scattering.

\subsection{Data to potential inversion for uncoupled situation}\label{dtopunco}

For clarity we suppress spin-related subscripts and begin by  
recasting Equation~\ref{second}, using Equation~\ref{first}, as 
\cite{early,ketal1,candm}:
\begin{equation}
\frac{\partial S_l}{\partial \alpha^{(p)}_n} =
 \frac{{\rm i} m }{\hbar^2 k}\int_0^{\infty} 
(\psi_l(r))^2 v^{(p)}_n(r) 
{\rm d}r. \label{third} \end{equation} 
We now introduce a conventional $\chi^2$ function:
\begin{equation}
\chi^2 = \sum^N_{k=1} \left(\frac{\sigma_k-\sigma_k^{\rm in}}
{\Delta \sigma_k^{\rm in}} \right)^2 +
\sum_n \sum^M_{k=1} \left(\frac{P_{kn}-P_{kn}^{\rm in}}
{\Delta P_{kn}^{\rm in}} \right)^2 \label{fourth}
\end{equation}
where $\sigma_k^{\rm in}$ and $P_{kn}^{\rm in}$ are the input
experimental values of cross sections and analyzing powers of type $n$
($\sigma$, ${\rm i}T_{11}$, etc.)
respectively. When fitting data for many energies at once, the
index $k$ indicates the angle and also the energy. 
Data re-normalising factors can  be introduced as an additional
contribution to Equation~\ref{fourth}.

We must now expand $\chi^2$ in terms of the $\alpha^{(p)}_n$. To do this
we first linearize the calculated cross sections and analyzing powers, by
expanding $\sigma_k$ (and $P_{kn}$)  about some current
point $\{ \alpha^{(p)}_n(i)    \}$ (see Ref.\cite{zuev}):
\begin{equation}
\sigma_k = \sigma_k(\alpha^{(p)}_n(i)) + \sum_{j,l} \left(
\frac{\partial \sigma_k}{\partial S_l(E_k)}\frac{\partial S_l(E_k)}
{\partial \alpha^{(p)}_n}\right)_{\alpha^{(p)}_n(i)}  
\Delta \alpha^{(p)}_n,
\label{new}
\end{equation}
 which applies at each iterative step $i =0, 1, 2,$\ldots and the
correction (to be determined) for the $n$-th amplitude is $\Delta \alpha_n^{(p)}
=\alpha^{(p)}_n - \alpha^{(p)}_n(i)$.
Equivalent relations are applied for the $P$'s.

Linear equations result from demanding that $\chi^2$ is locally
stationary with respect to variations in the potential coefficients
$\alpha_n^{(p)}$, i.e. the derivatives of $\chi^2$ with respect to the potential
components $\alpha_n^{(p)}$  must vanish.  Solving
these linear equations is straightforward for any reasonable number of
them and yields corrected values $\alpha^{(p)}_n(i)$. We then iterate
the whole procedure, with wave-functions $\psi_l$ in
Equation~\ref{third} calculated using the corrected potentials
from Equation~\ref{first}, until convergence is reached.  This algorithm almost
always converges very rapidly, in general diverging only when
highly inconsistent or erroneous data have been used or when the
iterative process involves a very unsuitable starting point.
Multi-energy ${\rm (data)}\rightarrow V$ inversion is thus reduced to 
the solution of simultaneous equations in a series of iterative steps.

\subsection{Generalisation to spin-1}
Spin-1 ${\rm (data)}\rightarrow V$ inversion is a natural generalisation of
the above formalism with $S_l$ replaced by $S^J_{l'l}$ and
Eq.~\ref{third} replaced by the analogous form derived from Eq.\ref{integ}.
It is shown in Ref.~\cite{prl} that the system does  indeed converge to a 
potential which fits the observables. 
\subsection{Evaluation of ambiguities for spin-1 ${\rm (data)}\rightarrow V$ 
inversion}\label{direct}
The tests of ${\rm (data)}\rightarrow V$ inversion must reflect the 
way it will be applied; this is rather different than for $S \rightarrow V$
inversion. With the latter, one often has quite precise $S$ calculated
from a theory, and one then seeks quite precise and subtle properties
of $V$, often relating to modifications of the theory. Inversion from 
measured observables is different because the data are generally far from
complete and will contain statistical and, possibly, systematic
errors. For this reason, we  must be less ambitious concerning the details
of the potential to be extracted. The test therefore
ask the following question: for a situation with few active partial waves,
how well-determined can we expect the potential to be? 

As in $S\rightarrow V$ inversion, one must never attempt to establish details 
of the potential for which the input data carries no information. We must
therefore apply the smallest possible inversion bases and accept approximate
solutions. The penalty for excessive inversion basis dimensionality
is the occurrence of spurious oscillatory features. In effect,
 at low energies where the data is incomplete and
featureless (reflecting the small number of partial waves),
the goal of  ${\rm (data)}\rightarrow V$ inversion
is to find the smoothest potential compatible with the data. IP inversion 
affords a level of control in this respect that is not possible with
other inversion procedures. 
 
The  test we describe is for low energy $\vec{\rm d}$ + $^4$He scattering.
The results will be useful for interpreting previous fits to experimental data 
for this system. The following observables, $\sigma$, ${\rm i}T_{11}$, 
$T_{20}$, $T_{21}$ and $T_{22}$, were calculated 
at laboratory  energies of 8, 9, 10, 11, 12 and 13 MeV using 
the same purely real, energy independent 
 potential used in Section~\ref{multi}.
Apart perhaps from the extremely strong `Dubovichenko-type' tensor
interaction, very strongly peaked at $r=0$, 
the general features of this potential are, we believe, 
similar to those of potentials  which fit actual experimental data.
 This energy range is somewhat above the broad 2$^+$
resonances and the region of strong mixing between the 1$^+$ channels. 
The observables were evaluated for the six energies 
over a range of 20$^0$ to 170$^0$ CM, at intervals of one degree, 
and Gaussian noise was added as follows. For $\sigma$, 1\% errors were imposed.
For ${\rm i}T_{11}$, the errors were 2\% of the maximum magnitude and 
for the three tensor observables, 5\% of the maximum magnitude.

We then applied ${\rm (data)}\rightarrow V$ inversion to
this multi-energy dataset, seeking  a single energy independent potential. 
Following Section~\ref{multi} and Ref.~\cite{prl}, 
the inversion bases for the Wigner and Majorana 
real central components consisted of three Gaussian functions. For the other
components there were just two Gaussians. As in  Section~\ref{multi},
the starting potential was zero in all but the Wigner real central and Wigner
real spin-orbit components. The results are shown in 
Figure 4 where we compare the known (`target') potential (solid lines),
the chosen starting potential of the iterative method (dash-dotted lines, two 
components only), and two inverted potentials, shown as dashed and dotted 
lines. The dashed lines show the potential found after a first
sequence of iterations and  correspond to $\chi^2/F = 15.473$
where $F$, the number of degrees of freedom, was $\sim 4500$. 
This number arises since we 
seek simultaneous fits  to five observables at 
six energies and 150 angles. The effective number of parameters 
is $\sim$twelve.  At this stage the reproduction of the larger components of 
the potential is fair, but the tensor terms are poor, with the even parity 
real tensor term being still almost zero.
The corresponding fit to the model data is indicated by the set of dashed lines
in Figure 5.  The fit is of a quality which would be widely regarded as quite
good when fitting experimental data, with only $T_{22}$, and perhaps 
$T_{20}$ around 120 degrees,  fitted poorly. The quality of fit to $T_{21}$ 
is remarkable in view of the very poor reproduction of the tensor interaction.

A subsequent further set of iterations led to an almost perfect fit with 
$\chi^2/F = 1.2155$. Figure 4 shows that the potential, dotted, 
fits all parts of the potential except at quite small radii. In particular,
the even parity real tensor is perfectly fitted for $r>1$ but not
fitted at all for $r<1$. This is in accord with arguments given in
Section~\ref{multi}. As expected from the values of $\chi^2/F$, the fits
to the 10 MeV dataset, shown as dotted lines in Figure 5, are essentially 
perfect, being scarcely visible over the angular range of
the artificial data. The same potential simultaneously fits   the
observables for the other five energies comparably well. We conclude
that we could not expect to establish  the various components of
the potential to a higher degree of accuracy than shown in Figure
4 by fitting available experimental data. It is very salutary
to see, in Figure 4, the profound change in the nature of the tensor 
interaction which follows the improvement of the fit revealed in Figure 5, 
comparing dashed and dotted lines. The intermediate inversion,
dashed lines in Figure 5, represents a fit of a quality which is
often deemed acceptible when fitting experimental data.
We note without further comment that the desirability of pursuing the best 
possible phenomenological fits is sometimes called into question. 

It should be noted that the computing time required on a modern
workstation to carry out the direct inversion of the data is very modest,
and certainly much less than required to carry out a
model independent optical model search, particularly one involving
odd and even parity $T_{\rm R}$ components and about 4500 degrees of
freedom. 

In Ref.\cite{prl58} we discussed the application of
direct inversion of data as a method for phase shift analysis. It is therefore
of interest to see the quality of fit to $S^J_{l'l}$ which corresponds
to the two fits shown in Figure 5. The top three panels 
of Figure 6 show the phase shifts corresponding to the 
$l=J-1,\,\, l=J,\,\, l=J+1$ diagonal components of $S$, and the
bottom  panel presents half the argument of the non-diagonal $S$.  
The solid lines show the known
potential, the dashed line is for the $\chi^2/N= 15.473$ fit and the solid
line is for the $\chi^2/N = 1.2155$ fit. For two of the panels, the solid
and dash-dot lines are nearly indistinguishable but they
are clearly distinguishable in the other two, suggesting that there are
limits to phase shift determination
even when over some 4000 data are fitted with $\chi^2/N = 1.2155$.

We conclude that direct inversion  is a
practical, reliable and efficient means of determining  a local
potential  which represents large, 
multi-energy datasets  including tensor observables.  The example 
presented here indicates the extent to which the
results obtained by this method are meaningful at low energies where
few partial waves are involved.

\section{Summary and conclusions; survey of possible applications} 
We have presented details of an inversion procedure which can be applied
both to spin-1 projectiles scattering from a spin-0 target nucleus and to 
spin-$\frac{1}{2}$ plus spin-$\frac{1}{2}$ particle scattering. The 
non-diagonal $S^J_{l'l}$ yield a non-diagonal potential containing a 
tensor term. To our knowledge, this is the first time this has been 
achieved, and opens up the possibility of wide range of other inversion 
scenarios ranging from other channel spin-1 cases (such as p + $^3$H 
scattering) to the inversion of $S$-sub-matrices of higher dimensionality.
There are many other capabilities inherent in the
underlying IP method: these include the possibility of inverting $S^J_{l'l}$ 
for several energies leading directly to an energy dependent potential,
including bound state energies within the input data,  and
the ability to handle cases where parity dependence must be allowed for. 

In this paper we have presented tests  of IP $S\rightarrow V$ spin-1 
inversion and evaluated its performance in `difficult' cases. 
We showed that when there are sufficient active partial waves, the procedure 
yields very accurate potentials even quite near the nuclear centre. Where, 
on the other hand, there are few partial waves available to define each potential
component, as is typical with light nuclei at low energies and where the
potential is parity dependent, it is still possible to extract the qualitative
features of a potential.

We also addressed the fact that the method is at present limited to 
$T_{\rm R}$ tensor interactions although it is quite probable that 
processes leading to $T_{\rm P}$ interactions are active. We showed that
$S^J_{l'l}$  arising from $T_{\rm P}$ interactions can be fitted
quite well with a $T_{\rm R}$ tensor interaction and that, moreover,
this does not lead to serious errors in the non-tensor components of
the potential. 

The IP inversion algorithm also forms the basis of a very efficient 
alternative way to find a  multi-component local potential which 
fits elastic scattering data, particularly for multi-energy datasets.
This is the direct (observable) $ \rightarrow V$
inversion procedure in which the IP $S \rightarrow V$ inversion is
embedded. This `direct inversion' can be applied to spin-1 projectiles. 
We examined the ambiguity problems which arise in a `difficult' 
(i.e. few partial waves, parity dependence) test case which is
relevant to the evaluation of an
analysis of low energy $\vec{\rm d}$ + $^4$He scattering, the subject of 
a recent~\cite{prl} and an extended future publication.  
Known potentials can be very well re-fitted, but it is clear that
the non-central terms cannot be well established at the nuclear centre. In the
course of performing this inversion test, it became
apparent  that fits of widely accepted quality lead to tensor potentials
which have nothing in common with those determined by pursuing `perfect fits'.

Finally, we remark that the method we have demonstrated here is certainly
not limited in usefulness to deuteron scattering. It would certainly
be worthwhile applying it the elastic scattering data for halo nuclei 
when these are of sufficiently substantial information content.

\section*{Acknowledgements} We are most grateful to the UK EPSRC for 
grants supporting S.G. Cooper, the Russian Foundation for Basic
Research (grant 97-02-17265) for financial assistance and to the Royal
Society (UK) for supporting a visit by V.I. Kukulin to the UK. We thank Jeff 
Tostevin for sending us Goddard's deuteron scattering code DDTP.

\setlength{\parindent}{0.0 in}

\newpage

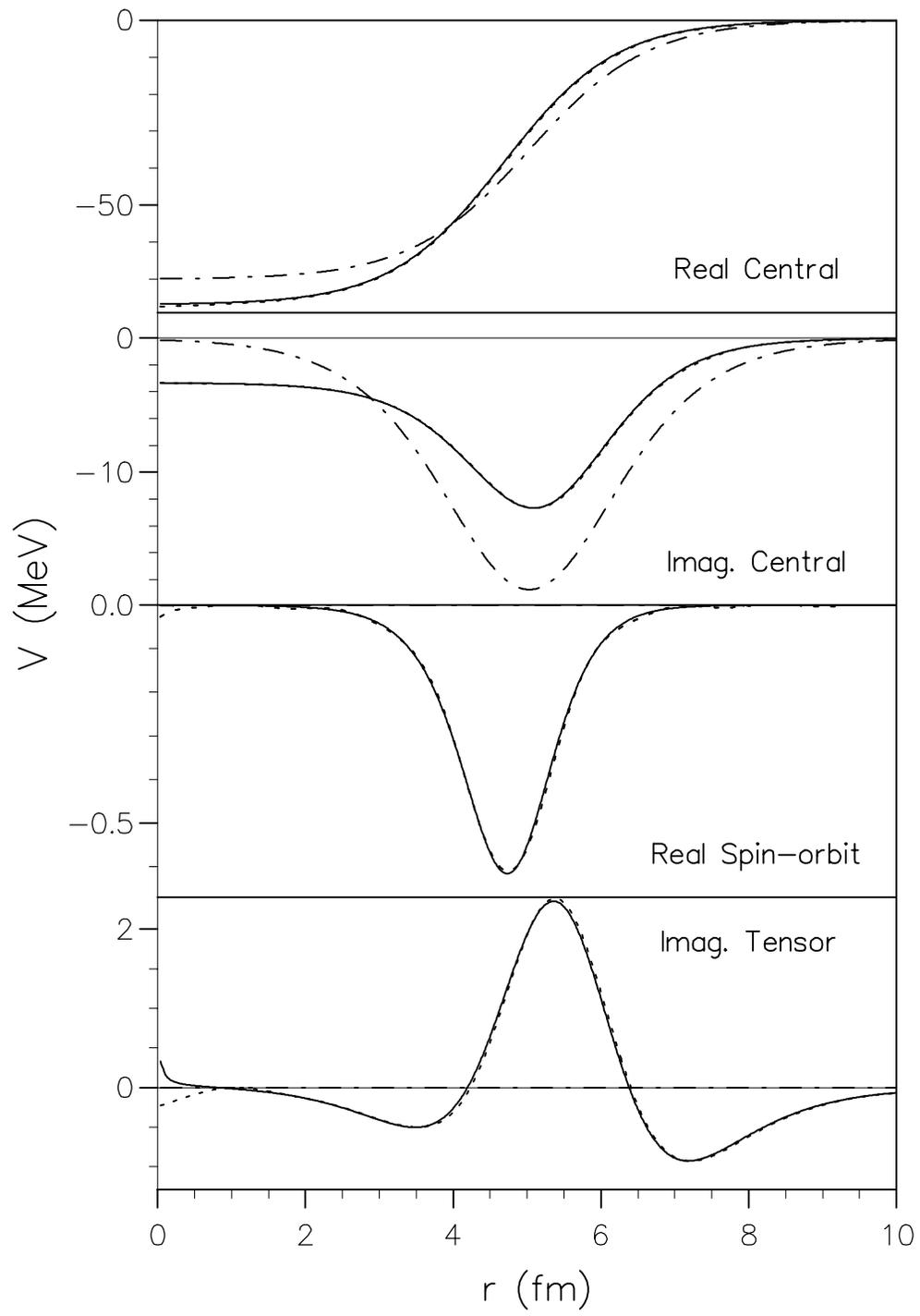
\begin{figure}
\caption{Potential for deuterons scattering from $^{58}$Ni at 56 MeV. From top,
the real central, imaginary central, real spin-orbit and imaginary tensor terms.
Solid lines represent the target (i.e. input) potential, dots the potential
found by inversion and the dash-dot the starting potential of the inversion.
The starting potential was zero for the real spin-orbit and the tensor terms.
There was no imaginary spin-orbit potential; see text regarding the real 
tensor term.}
\end{figure}

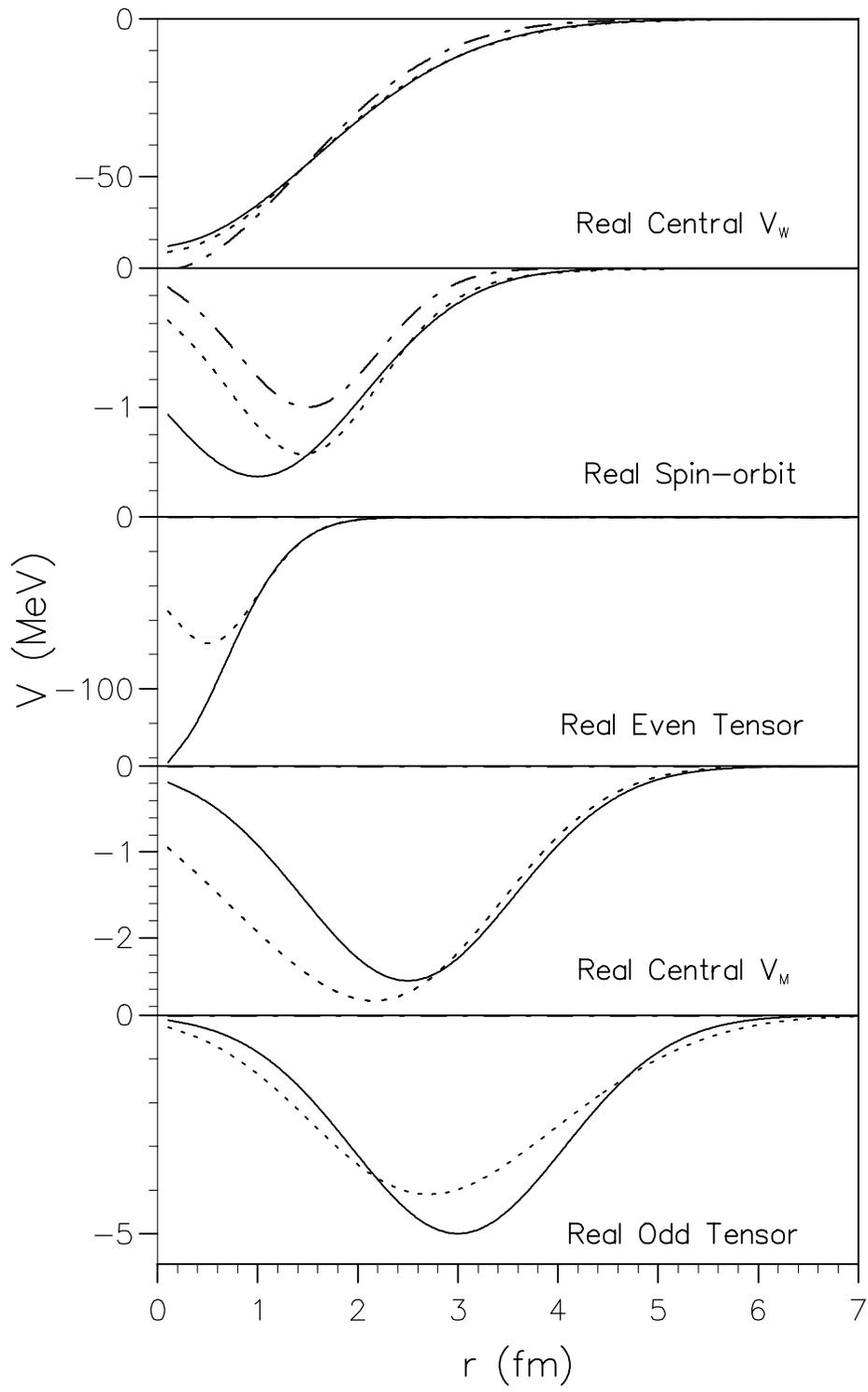
\begin{figure}
\caption{For deuterons scattering from $^4$He for energies from 8 -- 13 MeV, the
starting (dash-dot), inverted (dotted), and known, `target', (solid) potentials. 
From the top, the potential components are: Wigner real central, 
Wigner real spin-orbit, even parity real $T_{\rm R}$, Majorana real central 
and odd-parity real $T_{\rm R}$.}
\end{figure}

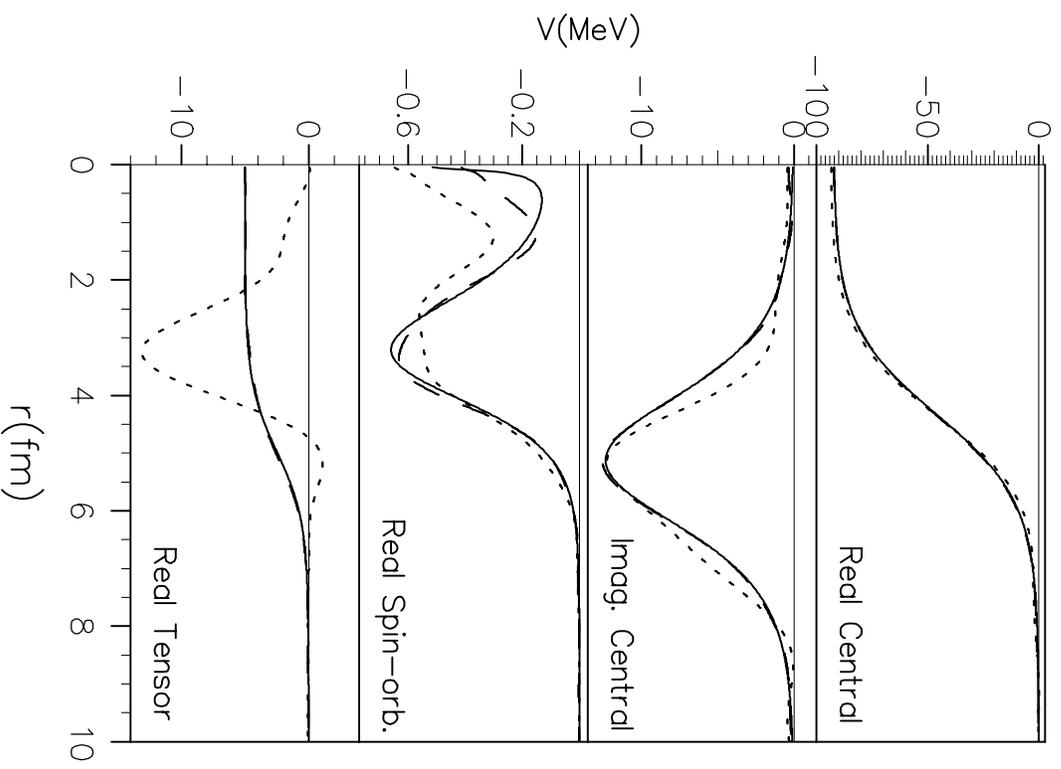
\begin{figure}
\caption{For deuterons scattering from $^{56}$Fe at a laboratory energy of 30 MeV,
comparing components (from top,
the real central, imaginary central,  real spin-orbit and real $T_{\rm R}$) 
of the potential of 
Goddard  (solid line) having a $T_{\rm R}$ term, the potential (dashes) 
found by inverting $S^J_{l'l}$ calculated from the solid line potential and
 the potential (dotted)  found by inverting $S^J_{l'l}$ calculated
from Goddard's `equivalent' $T_{\rm P}$-containing potential. The dashed and
solid lines are only clearly distinguishable for the smallest, i.e.\ spin-orbit,
component.}
\end{figure}

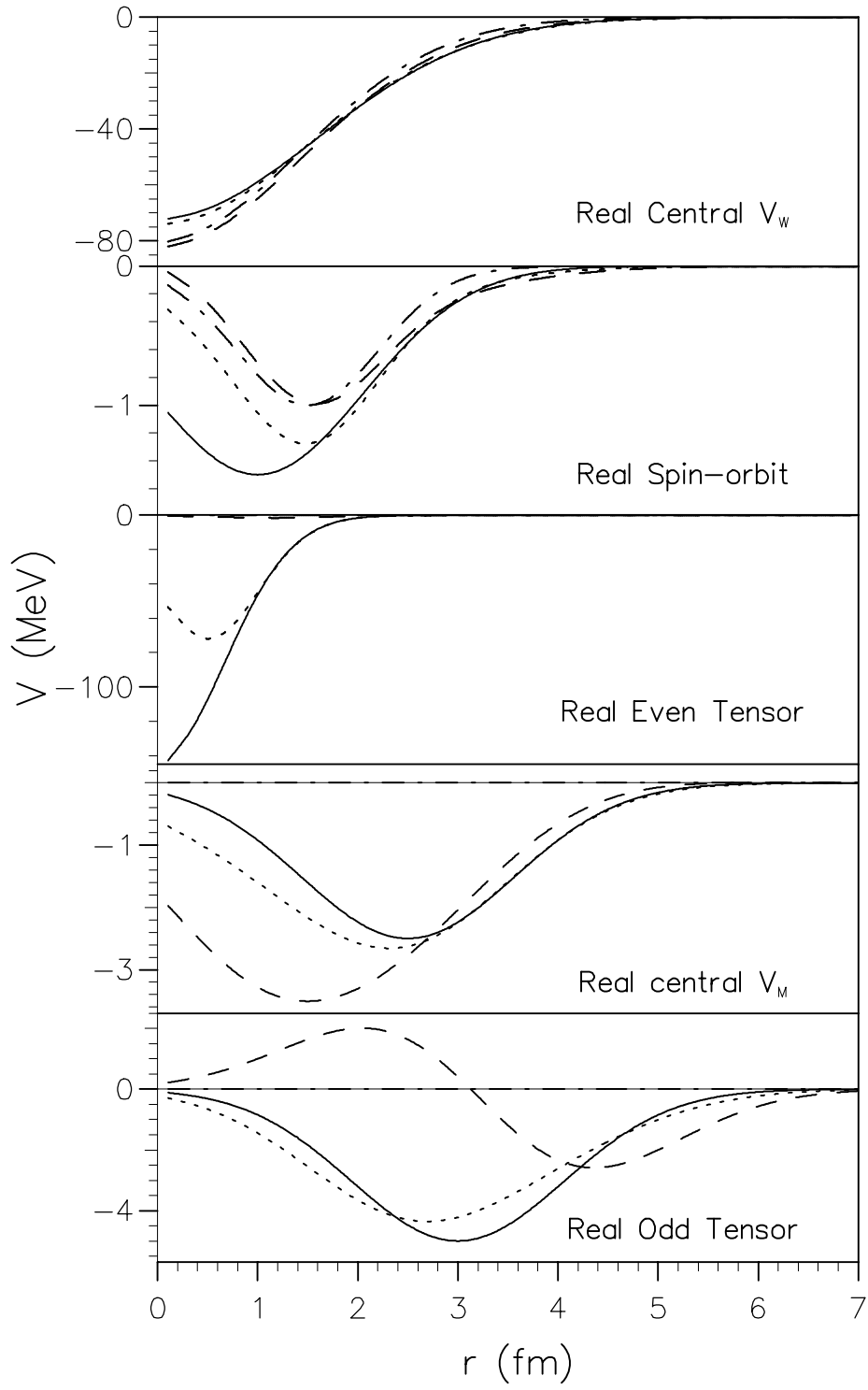
\begin{figure}
\caption{For deuterons scattering from $^4$He for energies from 8 -- 13 MeV, 
potentials found by direct inversion of artificial data incorporating 
Gaussian noise as described in the text. The starting potential, two 
components only, is shown as a dash-dotted line, the true (`target') potential is 
solid and the first stage ($\chi^2/N= 15.473$) and second stage 
($\chi^2/N = 1.2155$) inversion potentials are shown dashed and dotted 
respectively.}
\end{figure}

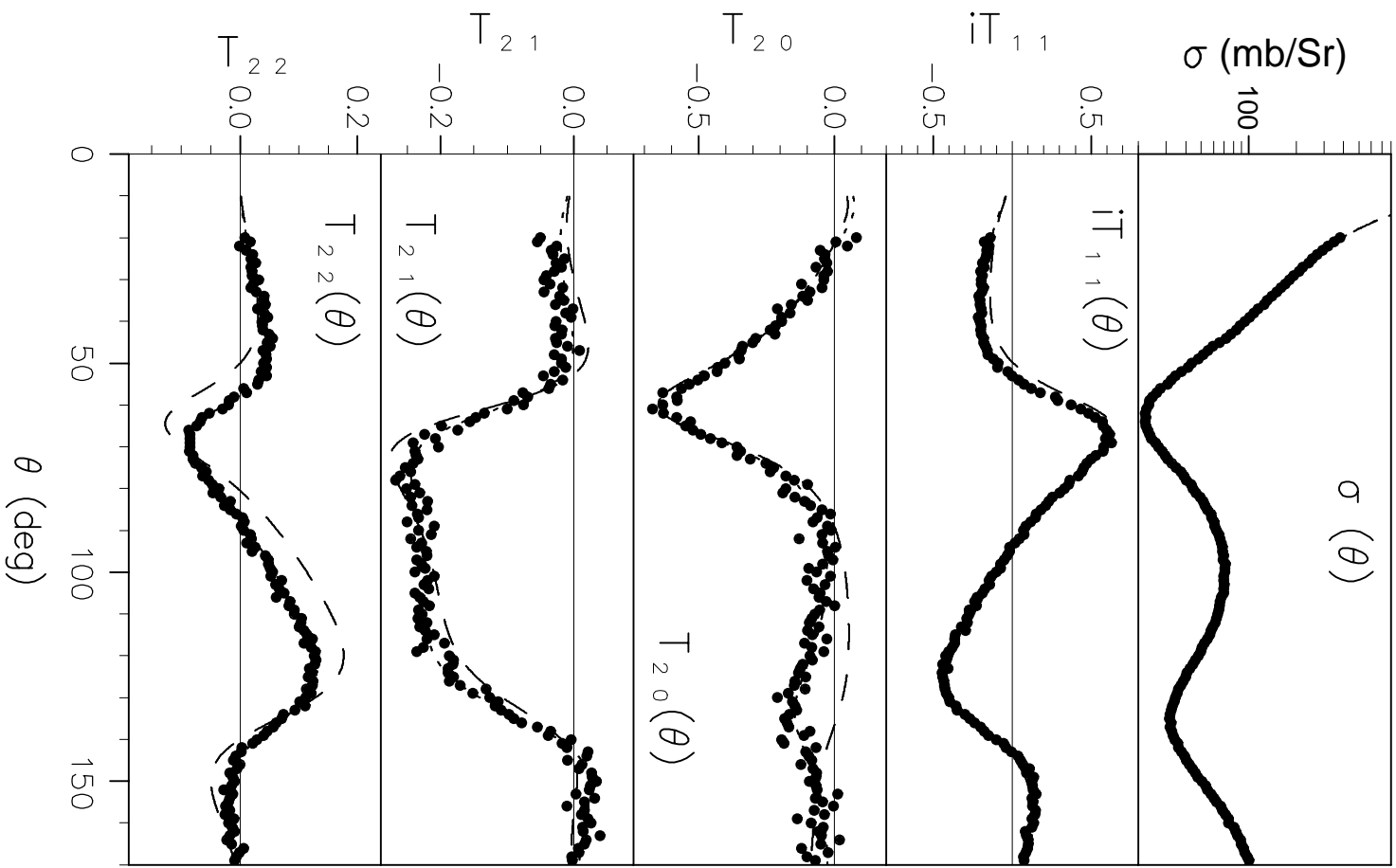
\begin{figure}
\caption{For deuterons scattering from $^4$He for energies from 8 -- 13 MeV,
the fits to the 10 MeV part of the six energy artificially generated datasets. 
The final fit with $\chi^2/N = 1.2155$ is the dotted line (visible in a few
places only) and the intermediate fit with $\chi^2/N= 15.473$ is shown dashed.}
\end{figure}

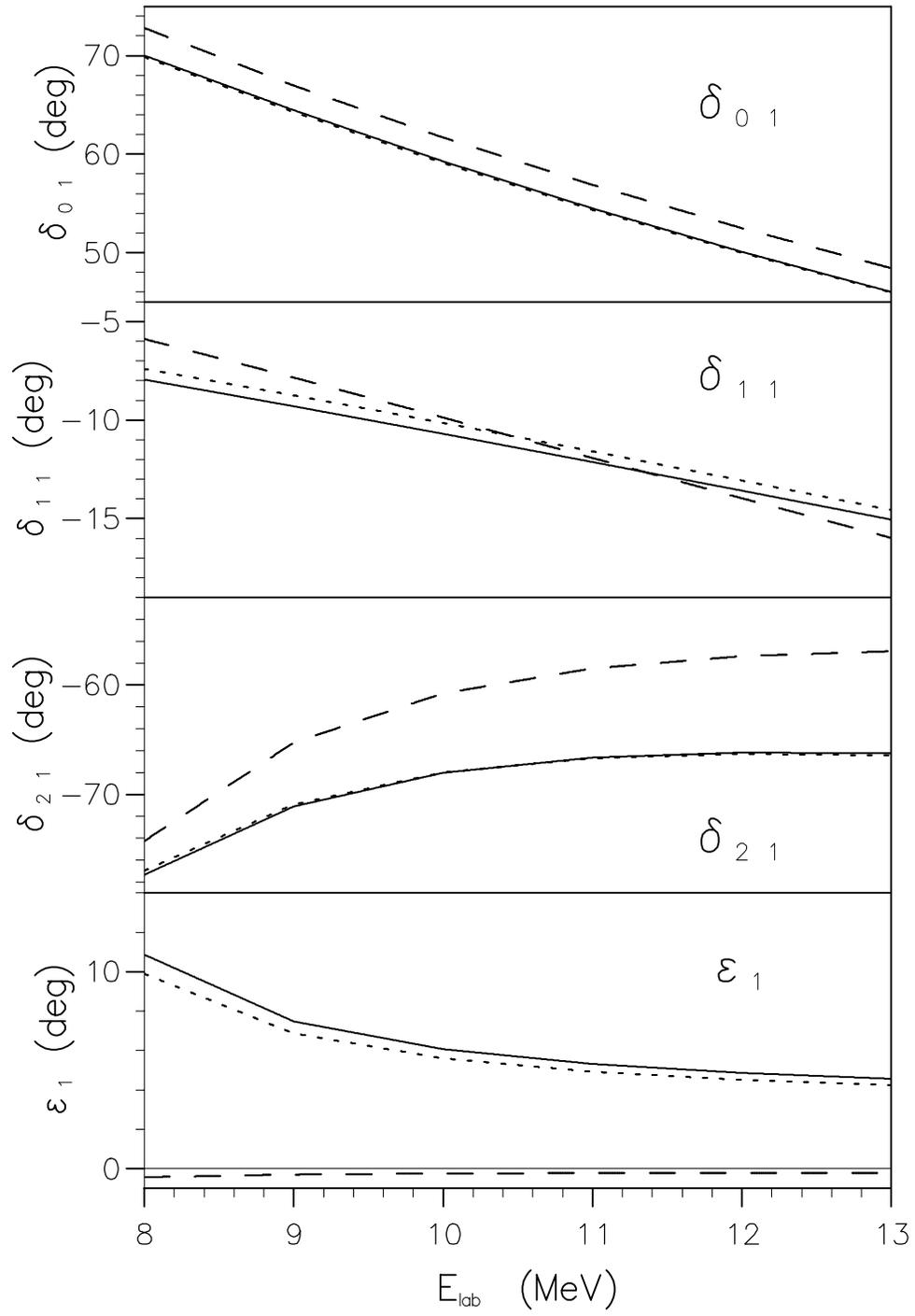
\begin{figure}
\caption{For deuterons scattering from $^4$He for energies from 8 -- 13 MeV,
fits to the diagonal (top three panels) and non-diagonal $S$-matrix.
The solid curves correspond to the known potential, the  
$S$-matrix for the $\chi^2/N= 15.473$ fit is dashed and for the
$\chi^2/N = 1.2155$ fit is dotted. As described in the text,
the four panels are essentially the (real) phase shifts. }
\end{figure}

\end{document}